\documentclass[12pt]{article}
\usepackage{graphicx}

\usepackage[numbers]{natbib}
\title{A semiclassical non-adiabatic theory  for elementary chemical reactions}

\author{S. Aubry \\
Laboratoire L\'eon Brillouin(CEA-CNRS)\\
CEA Saclay,
91191-Gif-sur-Yvette, France
\\serge.aubry91\@gmail.com}

\begin{document}
\maketitle

\small{\textbf{Abstract:} \textit{Electron Transfer (ET) reactions are modeled by the dynamics of  a quantum two-level system 
(representing the electronic state) coupled to a thermalized bath of classical harmonic oscillators (representing the nuclei degrees of freedom). 
Unlike for the standard Marcus theory, the complex amplitudes of the electronic state are chosen as reaction coordinates.
Then, the dynamical equations at non vanishing temperature become those of an effective Hamiltonian submitted to damping terms and their associated  Langevin random forces.
The advantage of this new formalism is to extend the original theory by taking into account both ionic and covalent interactions.
The standard theory is recovered only when covalent interactions are neglected. Increasing these covalent interactions from zero,
the energy barrier predicted by the standard theory first depresses,  next vanish (or almost vanish) and for stronger covalent interactions,
covalent bond formation takes place of ET.
In biochemistry, the standard Marcus theory often fails to explain the enzymatic reactions especially those with non Arrhenius behavior
which are  barrierless and also dissipate little heat. We claim that this improved theory should yield an interesting tool for understanding them. }

\textbf{keywords:} \textit{Chemical reactions, Electron Transfer, Covalence, Ionicity, Mixed Valence.}}







Chemical reactions are primarily changes of electronic states (associated with molecular and environmental reorganization) which appear either in radical ionization (redox) or 
in the forming/breaking of chemical bonds. They can be generally decomposed into sequences of elementary chemical reactions (ECR), each of them  corresponding
to a single transition between two different electronic states. The simplest example of ECR is an electron transfer
(ET) between a Donor and an Acceptor but it may also correspond more generally to exciton creation, exciton transfer etc.... 
The rate of chemical reactions often obeys the Arrhenius law which manifests the existence of
an energy barrier between the reactants and the products which has to be overcome under the effect of thermal fluctuations. 
These energy barriers are usually quite large compared to the room  temperature energy ($\approx 0.026 eV$ at $300 K$).
There are also   chemical reactions which do not obey the Arrhenius law (with a positive energy barrier).
This is the situation for  \textit{free radicals} with unpaired electrons which are often highly reactive
and generate covalent bonds.

\begin{figure}
    \centering
   \includegraphics[angle=270,width=
   \textwidth]{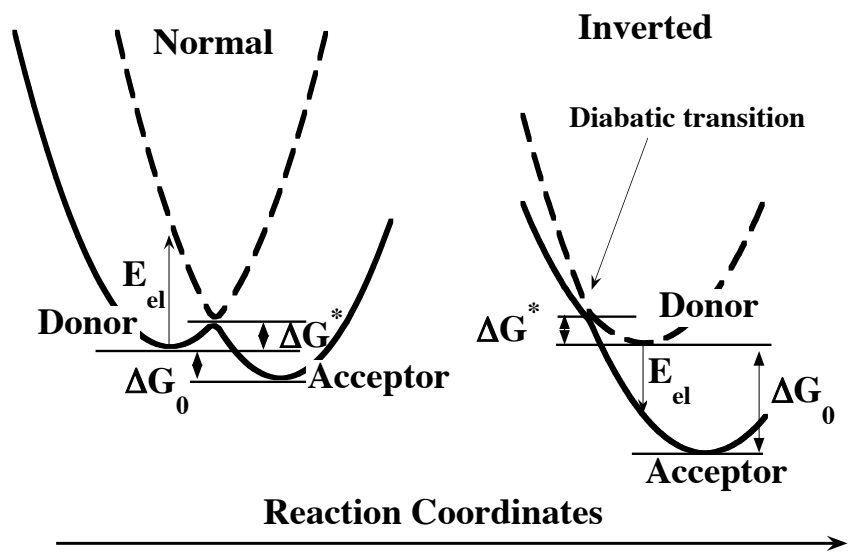}
    \caption{Standard Theory: The surfaces of energy versus Reaction Coordinates are  two intersecting paraboloids, one corresponding to
  the reactants and  one to the products. A small gap opening at intersection determines the lower energy (adiabatic) surface (thick line)
  and the upper energy (adiabatic) surface (dotted line).}
 \label{fig1}
\end{figure}

The standard theory for ET (redox) mostly due to \citet{Mar93}, considers the free energy of the whole system as a function
of the nuclei (reaction)  coordinates when the electron is on the Donor site (reactants) or  on the Acceptor site (products).
These functions are approximate as paraboloids schematically represented fig.\ref{fig1}.
There are two regimes called normal when at constant coordinates, the electronic excitation from Donor to Acceptor requires 
to absorb a positive energy $E_{el}$ and inverted when $E_{el}$ is negative.

The lowest point at the intersection  of these two surfaces determines the minimum free energy $\Delta G^{\star}$ to be provided to the system 
 for transferring the electron between Donor and Acceptor.  This energy barrier may be reached because of the thermal fluctuations of the nuclei 
 with a probability per unit time proportional to $e^{-\frac{\Delta G^{\star}}{k_BT}}$ which yields the main factor of the Arrhenius law. At this point 
 the two electronic states on Donor and Acceptor are degenerate so that ET may occur by  quantum tunneling. Actually, it is assumed there
 is an overlap between the two orbitals  necessary for allowing tunneling but which also raise the degeneracy and open a gap at the intersection 
 between the two   diabatic surfaces fig.\ref{fig1}. This overlap (and gap) is generally assumed to be small. Then, one obtains two non-intersecting adiabatic surfaces.
However, the real evolution of the electronic wave function is not considered as a continuous process. Instead of,  electron tunneling is  considered as a 
discontinuous jump which occur with some probability depending on  the time during which resonance lasts thus depending on the phonons (or vibrons)  and the temperature. 
The transition probability between the two diabatic surfaces $A(T)$ which contributes to the prefactor of  the Arrhenius law is empirically  
calculated from  the Landau-Zener model. In the normal regime,  ET does not require any diabatic transition while in the inverted regime  a diabatic transition 
from the upper  to the lower energy surface is necessary (see fig.\ref{fig1}). 
Current researches related to this problem of ET are still very active for example in the field of Organic Mixed Valence Compounds  \citet{review2,review}.

Our purpose is  to propose a semiclassical theory which describes the dynamics of ET (or ECR)   as a continuous process. The nuclei are still considered as classical particles while the dynamics of the electronic state is treated quantumly preserving possible phase coherence. Our model becomes equivalent to a quantum spin $1/2$ coupled to a  phonon bath (spin-Boson model). Early study of similar models were done for application to NMR \citet{Redfield}  and later  for chemical reactions  \citet{MM79,ST96,SWV08,Mil09}  but using  different approaches for example the density matrix theory.  However, our spin-boson model is different from those studied earlier because both the $z$ and $x$ spin components are coupled with the phonon bath
instead of the only $z$ component. Thus our model has the advantage to interpolate between a redox chemical reaction and a covalent bond formation which was never done before.
Otherwise, we use a simpler formalism for direct derivation of the quantum dynamical equations and then a  standard mean field approximation yields the semiclassical approximation.

We start from first principles and consider very generally, the global quantum  Hamiltonian $H = H_e(\mathbf{R}) + \sum_i \frac{\mathbf{P}_i^2}{2M_i}$ of our reacting system
consisting of many interacting electrons $\alpha$ with coordinates $\mathbf{r}= \{\mathbf{r}_{\alpha}\}$
also interacting with a collection of quantum nuclei $i$ with masses $M_i$ and coordinates $\mathbf{R}_i$. 
It  is  the sum of the Hamiltonian $H_e(\mathbf{R})$ of the whole system of electrons in the potential generated by the nuclei 
and of their kinetic energy operator  which depends on their  momenta operators $\mathbf{P}_i= \frac{\hbar}{i} \nabla_{\mathbf{R}_i}$. 
The standard Born-Oppenheimer (BO) approximation assumes that the global wavefunction has the form $\Psi(\mathbf{r},\mathbf{R},t)= \Phi(\mathbf{R},t) \psi_0(\mathbf{r},\mathbf{R})$
where $\psi_0(\mathbf{r},\mathbf{R})$ is the electronic groundstate  of Hamiltonian $H_e(\mathbf{R})$. Its eigen energy $E_0(\mathbf{R})$
becomes the interaction potential of the nuclei. Theories done in the framework of the BO approximation \citet{TST} consider that chemical species correspond to local minima of this potential energy surface and that chemical reactions are transitions between those minima induced by thermal fluctuations which help to overcome energy barriers (transition state).
These theories are valid when the other electronic states remain far  in energy  so that they are not involved.

We consider now an ET  which involves an electronic subspace $\mathcal{E}(\mathbf{R})$ spanned by two diabatic states  $\psi_D(\mathbf{r},\mathbf{R})$ and $\psi_A(\mathbf{r},\mathbf{R})$ real and orthogonal (LCAO representation). For a standard ET, the diabatic electronic state $D$ would be the initial state  with an electron on the Donor site and the final state $A$ those with this electron transferred on the Acceptor site. Thus, we assume that during the transition, the electronic state remains confined in this 2D subspace   $\mathcal{E}(\mathbf{R})$ that is  the global wave function  takes  the form $\Psi(\mathbf{r},\mathbf{R},t)= \Phi_D(\mathbf{R},t) \psi_D(\mathbf{r},\mathbf{R})+ \Phi_A(\mathbf{R},t)\psi_A(\mathbf{r},\mathbf{R})$.  All the other electronic states are supposed to remain far in energy so that their possible hybridization can be discarded. 

Integration over all the electronic variables $\mathbf{r}$ yields the nuclei Hamiltonian $<\Psi(\mathbf{r},\mathbf{R},t)|H_e(\mathbf{R})+ \sum_i \frac{\mathbf{P}_i^2}{2M_i}|\Psi(\mathbf{r},\mathbf{R},t)>_r  = \tilde{H}_e+\tilde{K}$ which  operates  in the two-components  wave function space $\Phi(\mathbf{R},t) = \left(\begin{array}{ c}
     \Phi_D(\mathbf{R},t)   \\
     \Phi_A(\mathbf{R},t)  
\end{array} \right)$. $\tilde{H}_e$ is a $2\times 2$ matrix which has the form $\tilde{H}_e= \left(\begin{array}{ cc}
E_D(\mathbf{R})       &   
\Lambda(\mathbf{R})\\
\Lambda(\mathbf{R} )   &  E_A(\mathbf{R})
\end{array} \right) $ only dependent on $\mathbf{R}$ while the projected kinetic energy  operator 
$\tilde{K}=\sum_{i,\alpha} \frac{P_{i,\alpha}^2}{2M_i}$ can  be expressed with the following overlap integrals
$a_{i,\alpha}^{n,m}(\mathbf{R}) =  \frac{1}{M_i}  \int \psi_n(\mathbf{r},\mathbf{R}) \frac{\partial  \psi_m(\mathbf{r},\mathbf{R})} {\partial R_{i,\alpha}}  d\mathbf{r} $ for $n,m=D$ or $A$. 
Orthonormalization implies  $\mathbf{a}^{D,D}(\mathbf{R})=0$ and $\mathbf{a}^{A,A}(\mathbf{R})=0$ and
$a_{i,\alpha}^{D,A}(\mathbf{R})=-a_{i,\alpha}^{A,D}(\mathbf{R})=\frac{1}{M_i}  \int \psi_D(\mathbf{r},\mathbf{R})  \frac{\partial \psi_A(\mathbf{r},\mathbf{R}}
{\partial R_{i,\alpha}}  d\mathbf{r}$  real. We define vector $\mathbf{A}(\mathbf{R})= \{a_{i,\alpha}^{D,A}(\mathbf{R})\}=\{-a_{i,\alpha}^{A,D}(\mathbf{R})\}$.
We also define the matrix elements for $n,m=D$ or $A$,
 $ w^{n,m}(\mathbf{R})=w^{m,n}(\mathbf{R}) =
\sum_i\frac{1}{2M_i} \int \nabla_{\mathbf{R}_i}\psi_n(\mathbf{r},\mathbf{R}).\nabla_{\mathbf{R}_i}\psi_m(\mathbf{r},\mathbf{R}) d\mathbf{r}$.
Then,  the projected kinetic operator $\sum_i \frac{\mathbf{P}_i^2}{2M_i}$ becomes the $2\times2$ matrix of operators
$ \tilde{K} =\left(\begin{array}{ cc}
\mathbf{K}_{DD}     &  \mathbf{K}_{DA} \\
\mathbf{K}_{AD} &  \mathbf{K}_{AA} 
\end{array} \right) $
where
$\mathbf{K}_{DD}= \sum_i \frac{\mathbf{P}_i^2}{2M_i} + \hbar^2  w^{D,D}(\mathbf{R}) $,
$ \mathbf{K}_{AA}= \sum_i \frac{\mathbf{P}_i^2}{2M_i} + \hbar^2  w^{A,A}(\mathbf{R})$,
$\mathbf{K}_{DA}= -\frac{i \hbar }{2}(\mathbf{A}.\mathbf{P}+\mathbf{P}.\mathbf{A}) 
+\frac{\hbar^2}{2} \mathbf{\nabla}. \mathbf{A} +\hbar^2 w^{D,A}(\mathbf{R})=\mathbf{K}_{AD}^{\star}$.
Using the base of Pauli matrices  $\sigma^{x}, \sigma^{y}, \sigma^{z}$
(with standard commutation relations
$[ \sigma^x,\sigma^y] = 2i \sigma^z$, $[ \sigma^y,\sigma^z] = 2i \sigma^x$, $[ \sigma^z,\sigma^x] = 2i \sigma^y$),
the fully quantum Hamiltonian appears a collection of quantum nuclei coupled to a single quantum spin $1/2$
\begin{eqnarray}
\tilde{H}&=&\sum_i \frac{\mathbf{P}_i^2}{2M_i}  +\mathcal{V}_{ph}(\mathbf{R}) \nonumber \\
&+& \tilde{\Lambda}(\mathbf{R}) \sigma^x+  \Pi(\mathbf{R}, \mathbf{P})  \sigma^y +  \mathcal{W}(\mathbf{R}) \sigma^z
 \label{hamspin}
 \end{eqnarray}
where $\mathcal{V}_{ph}(\mathbf{R}) = \frac{1}{2} (E_D(\mathbf{R})+E_A(\mathbf{R}))
+\frac{\hbar^2}{2} (w^{D,D}(\mathbf{R})+w^{A,A}(\mathbf{R}))$,
$\mathcal{W}(\mathbf{R})= \frac{1}{2} (E_D(\mathbf{R})-E_A(\mathbf{R}))
+\frac{\hbar^2}{2} (w^{D,D}(\mathbf{R})-w^{A,A}(\mathbf{R}))$,
$\tilde{\Lambda}(\mathbf{R}) = \Lambda(\mathbf{R}) +\frac{\hbar^2}{2} \mathbf{\nabla}. \mathbf{A}(\mathbf{R}) +\hbar^2 w^{D,A}(\mathbf{R})$ and
 $\Pi(\mathbf{R}, \mathbf{P}) = \frac{ \hbar }{2}\left(\mathbf{A}(\mathbf{R}).\mathbf{P}+\mathbf{P}.\mathbf{A}(\mathbf{R}) \right)$.

For going further, it is now convenient to assume  that  1)  Potential  $\mathcal{V}_{ph}(\mathbf{R})$ is quadratic 
and then choosing the origin of the nuclei coordinates as well as the origin of the energies  at its minimum, the nuclei
potential takes the form
$\mathcal{V}_{ph}(\mathbf{R}) = \frac{1}{2} \mathbf{R}.\overline{\overline{\mathbf{M}}}. \mathbf{R} $
where $\overline{\overline{\mathbf{M}}}$ is a positive elasticity matrix.
2) We assume a linear behavior for  the spin coefficients  of $\sigma^z$ (charge coupling), $\mathcal{W}(\mathbf{R}) =  \mathcal{W}(\mathbf{0}) +  \mathbf{\nabla}.\mathcal{W}(\mathbf{0}) .\mathbf{R}$,
of $\sigma^x$ (covalent coupling) $\tilde{\Lambda}(\mathbf{R}) = \tilde{\Lambda}(\mathbf{0}) + \mathbf{\nabla}. \tilde{\Lambda}(\mathbf{0}).\mathbf{R}$ and
of $\sigma^y$ $\Pi(\mathbf{R}, \mathbf{P})= \hbar \mathbf{A}(\mathbf{0}).\mathbf{P}$.
We also assume $\mathbf{A}(\mathbf{0})=0$
for simplicity (though it would not be a big deal to conserve this coupling with $\sigma^y$)   \footnote{Despite the use of bases of diabatic states is ubiquitous  for understanding electronic structures and dynamics in physics and chemistry, they have no strict definition as proven in ref. \citet{Diabpict}.  A good criteria to find them, is that
 they depend smoothly on the nuclei coordinates $\mathbf{R}$ with overlaps  $\mathbf{A}(\mathbf{0})$  as small as possible. Actually this assumption is necessary to justify the above low order expansion.}.
Then, it is convenient to use the base of normal modes obtained by diagonalization of matrix $\overline{\overline{\mathbf{M}}}$.

Hamiltonian (\ref{hamspin}) becomes that of a single quantum spin $1/2$  submitted to an external field  $(\epsilon_x,0,\epsilon_z)$ and
linearly coupled to a collection of quantum normal modes $n$ (harmonic oscillators with unit mass  and frequency $\omega_n$) 
by  constants $k_n^x,0,k_n^z$.
\begin{equation}
\tilde{H}= \sum_n \left( \frac{1}{2} \left(p_n^2 + \omega_n^2 q_n^2 \right) +k_n^x q_n \sigma^x +k_n^z q_n \sigma^z \right) + \epsilon_x \sigma^x + \epsilon_z \sigma^z
\label{hamtot4}
\end{equation}
Path integral methods were used for studying this model in the fully quantum case in the case  for $\epsilon_z=0$ and with no transverse coupling $k_n^x=0$ ( see \citet{Leg86} and references therein). It is nevertheless simpler and sufficient in our context to use a mean field  approximation where the coupling terms $q_n \sigma^z$  (and similarly for  $q_n \sigma^x$) are replaced by
$q_n \bar{\sigma}^z +\bar{q}_n \sigma^z -\bar{q}_n  \bar{\sigma}^z $  ( the expected value of an operator $a$ is $\bar{a}=< \Psi(t)|a|\Psi(t)>$ with $|\Psi(t)>$ the global wave function at time $t$). Thus, we neglect the fluctuation operators $(q_n -\bar{q}_n)( \sigma^z-\bar{\sigma}^z)$. This mean field approximation turns out to be equivalent to the standard classical approximation. It is valid when the relevant nuclei displacements generated by the molecular reorganization during ET are 
much larger than their quantum fluctuations  $<(q_n -\bar{q}_n)^2>^{1/2}$ which is often true in real situations.

The Ehrenfest equation $i \hbar \dot{\bar{a}}(t)=\overline{[\tilde{H},a]}$ provides closed equations for
the time derivatives of  $\{\bar{p}_n,\bar{q}_n\}$ and $\bar{\sigma}^x,\bar{\sigma}^y,\bar{\sigma}^z$ which do not involve any other variables.  They correspond to those 
  of a quantum spin $\mathbf{\sigma}$ submitted to the time dependent classical field
with components $(\epsilon_x+ \sum_n k_n^x \bar{q}_n(t),0,\epsilon_z +\sum_n k_n^z \bar{q}_n(t))$
and to classical oscillators $n$ submitted to external time dependent forces $f_n(t)=k_n^x \bar{\sigma}^x(t)+k_n^z \bar{\sigma}^z(t)$.
It is convenient to describe this spin $\varphi_D(t) | \uparrow > + \varphi_A(t) |\downarrow>$ with its two complex coordinates $\varphi_D(t)$ and $\varphi_A(t)$  
 in the eigenbase of $\sigma^z$ Donor-Acceptor (fulfilling the normalisation condition $|\varphi_D|^2+|\varphi_A|^2=1$).

Since  the nuclei variables obey linear equations,  the general solution $\bar{q}_n(t)$ of each linear oscillator can be explicitly calculated as the sum of a function of the external force $f_n(t)$
and a solution of the free oscillator  chosen randomly according to the Boltzman statistics  \citet{AK03,Aub07}.
It comes out after substitution in the spin equations that the dynamics of the electronic (or spin) components is described by two equations
corresponding to those of a Hamiltonian system where $ (\varphi_D^{\star}, i \hbar \varphi_D)$ and $( \varphi_A^{\star}, i \hbar \varphi_A)$
are pairs of conjugate variables but also submitted to dissipative and random forces
\begin{eqnarray}
 i \hbar \dot{\varphi}_D &=& \frac{\partial H_{eff}}{\partial \varphi_D^{\star}} + \zeta^z(t) \varphi_D + \zeta^x(t) \varphi_A \nonumber \\
&+&\left(  \int_0^t \left(\Gamma_{zz}(t-\tau) \dot{Z}(\tau) + \Gamma_{xz}(t-\tau) \dot{X}(\tau)\right) d\tau\right) \varphi_D \nonumber \\
&+& \left(  \int_0^t \left(\Gamma_{xz}(t-\tau) \dot{Z}(\tau) + \Gamma_{xx}(t-\tau) \dot{X}(\tau)\right) d\tau\right) \varphi_A \label{neweq1} \\
 i \hbar \dot{\varphi}_A &=& \frac{\partial H_{eff}}{\partial \varphi_A^{\star}} - \zeta^z(t) \varphi_A + \zeta^x(t) \varphi_D \nonumber \\
&-&\left(  \int_0^t \left(\Gamma_{zz}(t-\tau) \dot{Z}(\tau) + \Gamma_{xz}(t-\tau) \dot{X}(\tau)\right) d\tau\right) \varphi_A \nonumber \\
&+& \left(  \int_0^t \left(\Gamma_{xz}(t-\tau) \dot{Z}(\tau) + \Gamma_{xx}(t-\tau) \dot{X}(\tau)\right) d\tau\right) \varphi_D 
\label{neweq2}
\end{eqnarray}

The effective Hamiltonian in \ref{neweq2} is
\begin{eqnarray}
H_{eff} (\varphi_D,\varphi_A)&=& \epsilon_x X + \epsilon_z Z \nonumber\\ - \frac{1}{2} \Gamma_{xx}(0) X^2 &-& \Gamma_{xz}(0) XZ - \frac{1}{2}  \Gamma_{zz}(0) Z^2 
 \label{generham}
\end{eqnarray}
where $Z=\bar{\sigma}^z=  |\varphi_D|^2-|\varphi_A|^2$ and $X=\bar{\sigma}^x=\varphi_D^{\star}\varphi_A+\varphi_D\varphi_A^{\star}$.
$H_{eff}$  is nothing but the energy minimum with respect to all the nuclei coordinates of the whole system 
at fixed electronic amplitudes $\varphi_D$,$\varphi_A$  \footnote{A similar Hamiltonian was introduced phenomenologically in \citet{AK03} 
 without microscopic justifications but  the form which was used, was not correct because we  omitted covalent terms
  and artificially introduced extra nonlinear capacitive terms.}.
  
The memory (or damping) functions in \ref{neweq2} are defined as
\begin{eqnarray}
\Gamma_{xx}(t) &=& \sum_n \frac{(k_n^x)^2}{\omega_n^2} \cos \omega_n t = \int \tilde{\Gamma}_{xx}(\omega)   \cos \omega t ~d\omega \nonumber \\
\Gamma_{zz}(t) &=&\sum_n \frac{(k_n^z)^2}{\omega_n^2} \cos \omega_n t = \int \tilde{\Gamma}_{zz}(\omega)  \cos \omega t ~d\omega \nonumber \\
\Gamma_{xz}(t) &=& \Gamma_{zx}(t)=\sum_n \frac{k_n^z k_n^x}{\omega_n^2} \cos \omega_n t= \int \tilde{\Gamma}_{xz}(\omega)  \cos \omega t ~d\omega
\label{memf} 
\end{eqnarray}
They fulfill $0 \leq \Gamma_{zz}(0)$, $0 \leq \Gamma_{xx}(0)$ and $ \Gamma_{zx}^2(0) \leq  \Gamma_{zz}(0) \Gamma_{xx}(0)$.
Assuming a large system with many classical oscillators, these functions may be assumed to be smooth and to vanish at large time 
as for a standard classical Langevin bath. Then,  the absence of random forces $\zeta_{\alpha}(t)$,  this effective Hamiltonian  $H_{eff}$ necessarily decays in time
(essentially because of   nonadiabaticity) \citet{Aub07}.

Temperature appears  in eqs.(\ref{neweq1},\ref{neweq2}) through the gaussian random forces  with correlations fulfilling the Langevin conditions at temperature $T$
\begin{eqnarray}
<\zeta^z(t)\zeta^z(t+\tau)>_t &=& k_B T  \Gamma_{zz}(\tau) \nonumber  \\
<\zeta^x(t)\zeta^x(t+\tau)>_t &=& k_B T  \Gamma_{xx}(\tau)     \nonumber \\
<\zeta^x(t)\zeta^z(t+\tau)>_t &=& k_B T  \Gamma_{xz}(\tau) \label{Langevinrel}
\end{eqnarray}

The Fourier transforms of the memory functions (\ref{memf}) have to vanish at large 
frequency for $\omega>\omega_c$ where $\omega_c$ is the maximum phonon frequency. Then, far from electronic resonances $\Delta_{el} >> \hbar \omega_c$,
the characteristic frequency of the electronic dynamics is much beyond $\omega_c$ so that the effect of damping disappears. 
Eqs.\ref{Langevinrel} also show that the random forces  have a spectrum below $\omega_c$ with slow variations at the scale of the electron dynamics. 
We are in the adiabatic regime where the BO approximation is  valid. 
When the random forces bring the electronic system near resonance, the validity of the BO approximation breaks down
and  nonadiabaticity manifests by energy dissipation into the phonon bath.

\begin{figure} 
    \centering
      \includegraphics[width=   0.45\textwidth]{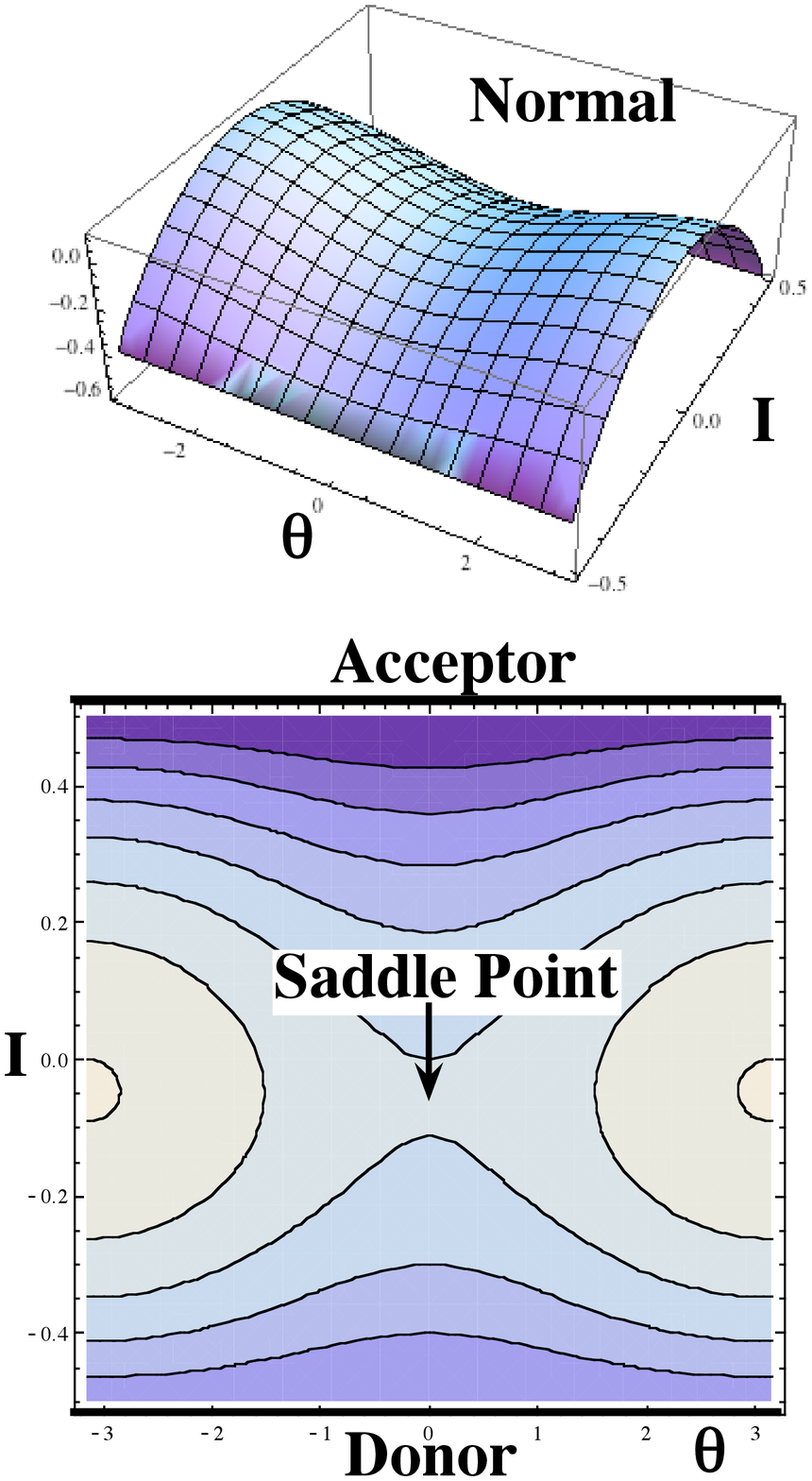}
        \includegraphics[width= 0.45 \textwidth]{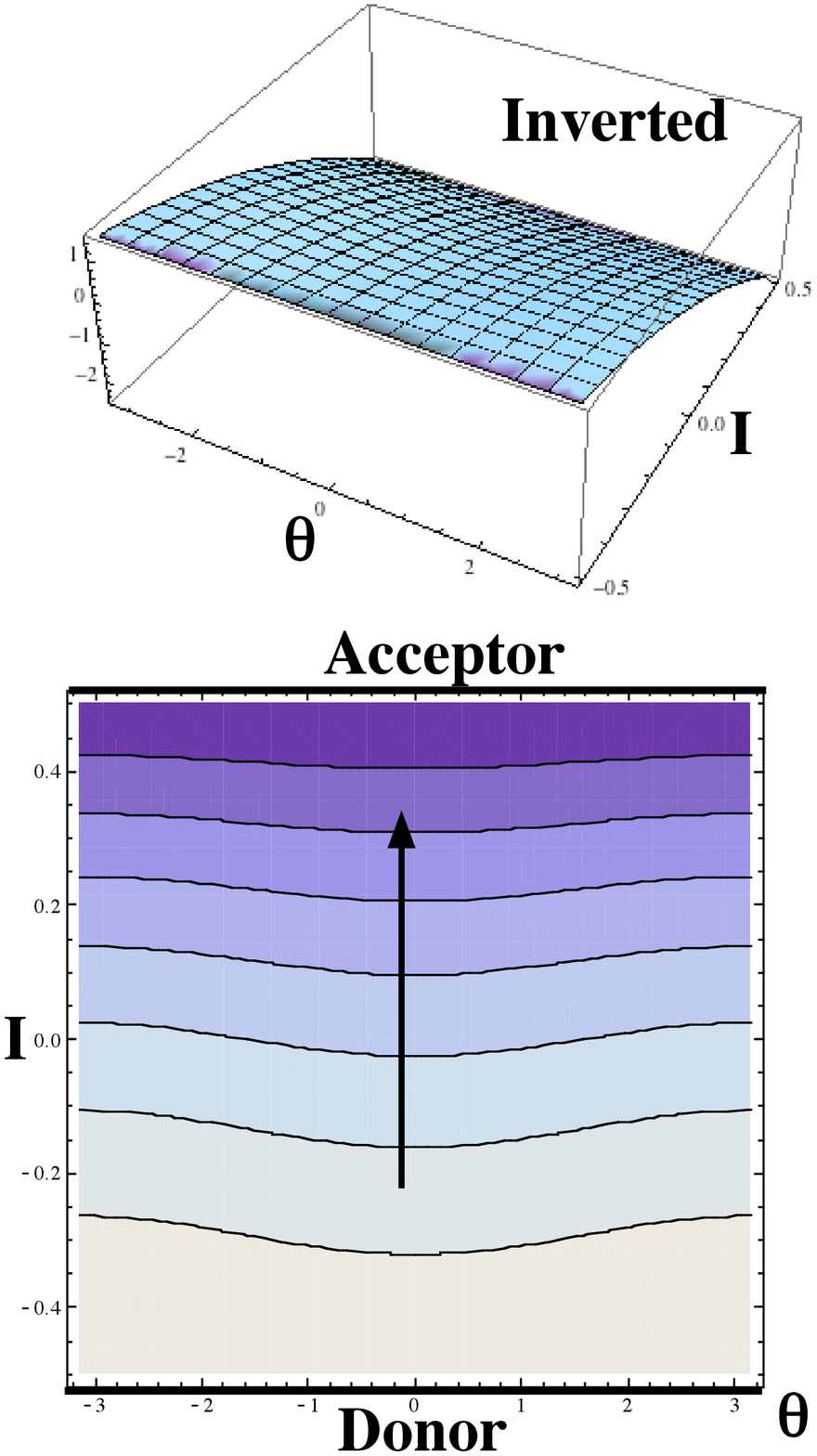}
    \caption{ Two examples of 3D plots of the the energy landscape $H_{eff}(I,\theta)$  in the normal (left) and the inverted regime (right) of the Marcus theory with below their corresponding contour plots (the sphere $I,\theta$ is represented within the Mercator projection where the poles are single points  appearing as the thick lines).
Covalent interactions ($\Gamma_{xx}(0)=\Gamma_{xz}(0)=0$) vanish and  $\Gamma_{zz}(0)=1.,\epsilon_z=0.1,\epsilon_x=-0.1$ (left)
and $\Gamma_{zz}(0)=1., \epsilon_z=1, \epsilon_x=-0.1$ (right). }
 \label{figsur1}
\end{figure}

\begin{figure} 
    \centering
      \includegraphics[width=
      0.45 \textwidth]{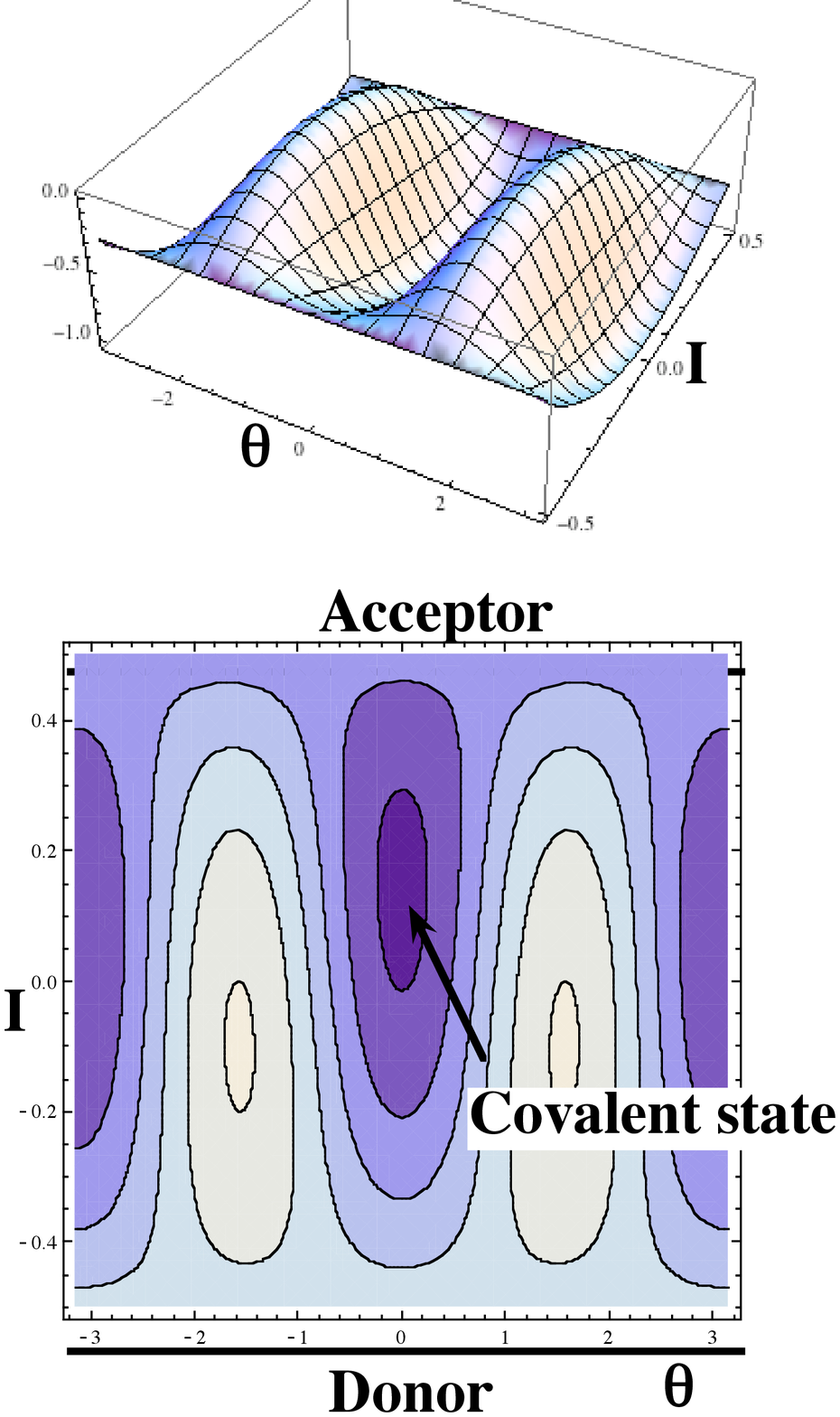}
        \includegraphics[width=
        0.45 \textwidth]{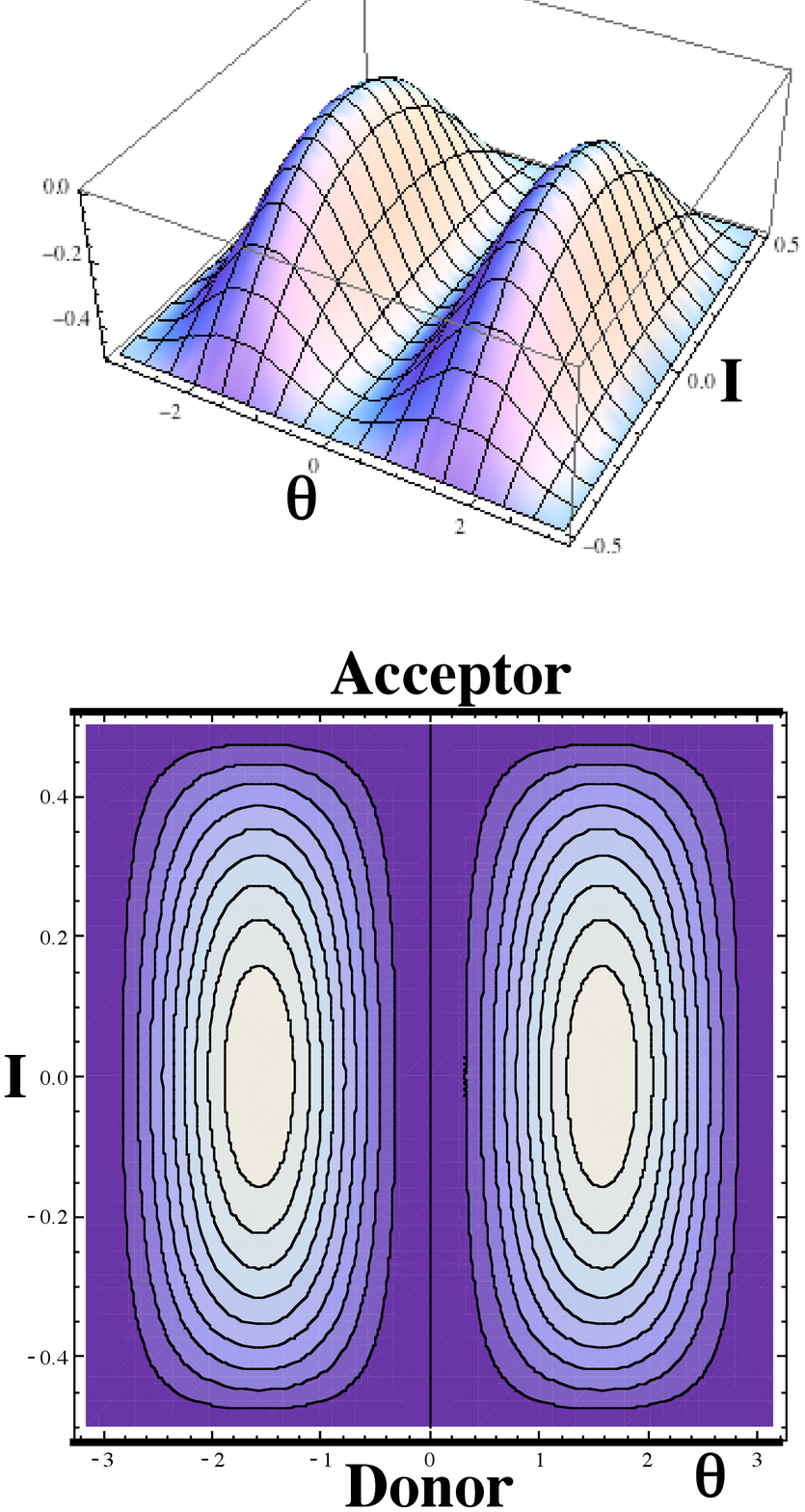}
    \caption{Same as fig.\ref{figsur1} but for some examples with covalent interactions. 
   On the left  side $\epsilon_z=0.2,\epsilon_x=-0.01,\Gamma_{zz}(0)=1,\Gamma_{xx}(0)=2.$ and $\Gamma_{xz}(0)=-0.1$.
correspond to a situation where the final state is a covalent state.
On the right side : $\epsilon_z=\epsilon_x=0$, $\Gamma_{xx}(0)=\Gamma_{zz}(0)=1.$ and $\Gamma_{xz}(0)=0$) 
  correspond to an ideally isoenergetic and barrierless situation.}
 \label{figsur2}
\end{figure}

Defining new conjugate variables $I_D,\theta_D,I_A,\theta_A$ by $\varphi_D=\sqrt{I_D}e^{-i\theta_D}$ and $\varphi_A=\sqrt{I_A}e^{-i\theta_A}$, and
next the conjugate variables $I=(I_A-I_D)/2= -Z/2$ and $\theta=\theta_A-\theta_D$,
this effective Hamiltonian becomes only a function $-\frac{1}{2}\leq I \leq \frac{1}{2},\theta$ mod $2\pi$
on a sphere with poles $I=\pm \frac{1}{2}$ and where
$\theta$ corresponds to the longitude and $\phi$ defined as $\sin \phi = 2*I$ to the latitude. 
Then
\begin{eqnarray}
H_{eff}&=&  -2 \epsilon_z I+ \epsilon_x \sqrt{1-4I^2}\cos \theta-2 \Gamma_{zz}(0) I^2\nonumber \\
&+&2 \Gamma_{xz}(0)  I  \sqrt{1-4I^2}\cos \theta  - \frac{1}{2} \Gamma_{xx}(0) (1-4I^2) \cos^2 \theta 
\label{effham}
\end{eqnarray}
represents the true energy surface on which the system evolves during ET. At zero temperature (0K) and assuming
the damping terms vanishes, the dynamical equations would be those of an integrable Hamiltonian system on a sphere.
All trajectories would be periodic on closed orbits defined by a constant  energy $H_{eff}(I,\theta)$.
Figs.\ref{figsur1} and \ref{figsur2} show both 3D and contour plots for several examples.

We assume that the (initial) transfer integral $\epsilon_x$ is small as in the standard theory and that
our system is  initially at the Donor pole $I=-1/2$. The Acceptor pole
corresponds to $I=+1/2$. When there are no covalent interactions ($\Gamma_{xx}(0)=\Gamma_{xz}(0)=0$), our theory
is nothing but a different representation of the standard Marcus theory except that we now have
explicit dynamical equations which intrinsically describe  the diabatic transitions through the damping terms
(without needing the Landau-Zeener model). The two Marcus energy surfaces are the paraboloids obtained from Hamiltonian (\ref{hamtot4})
where $\sigma^x=0,p_n=0$ and  $\sigma^z=+1$ (electron on Donor) or   $\sigma^z=-1$ (electron on Acceptor).
The reaction energy is $\Delta G_0=-2  \epsilon_z$ and $\lambda=2  \Gamma_{zz}(0) $ is the so called reorganization energy
known in the litterature. Then the barrier energy takes the standard form 
\begin{equation}
\Delta G^{\star} = \frac{(\lambda+\Delta G_0)^2}{4\lambda}
\label{enb}
\end{equation}
The electronic excitation energy is $E_{el}= 2 (\Gamma_{zz}(0)-\epsilon_z)=\Delta G_0+\lambda$.

In the normal regime, when $E_{el}>0$, the energy surface (see fig.\ref{figsur1} left) exhibits a saddle point
which corresponds to the top of the energy barrier between the donor and acceptor state. At the inversion point where $ E_{el}=0$, the saddle point 
and the two maxima on the sphere of $H_{eff}(I,\theta)$ merge with the  minimum near the pole $I=-1/2$ which thus become a single maximum
so that  in the inverted regime $E_{el}<0$, there is no more energy barrier (see fig.\ref{figsur1} right).
In both regimes, the two poles on the sphere Donor and Acceptor are surrounded by periodic and stable orbits with  frequency $\omega_{el}$
obtained by linearizing the equations (\ref{neweq1},\ref{neweq2}) which turns out to be related to $|E_{el}|=\hbar \omega_{el}$. 
When this electronic frequency $\omega_{el}$  is beyond the phonon spectrum that is $|E_{el}|>>\hbar \omega_c$,  the damping terms have negligible
dissipative effect in eqs.\ref{neweq1},\ref{neweq2} so that the poles are practically stable at 0K. At non vanishing temperature, 
the thermal forces  in eqs.\ref{neweq1},\ref{neweq2}, generate adiabatic random fluctuations of the electronic frequency $|E_{el}|$. 
If  these fluctuations bring $|E_{el}|$  in the phonon range  below $\hbar \omega_c$,  the damping terms in eqs.\ref{neweq1},\ref{neweq2} become efficient. 
The system is no more adiabatic and then  ET may occur with some probability. 
We have already shown in \citet{Aub07} that reaching this resonance is equivalent to reach the intersection between the two paraboloids
so that we recover a standard Arrhenius law (despite there is apparently no energy barrier for $H_{eff}$).
On contrary, in the inverted regime but near the inversion point  where  $E_{el} <\hbar \omega_c$, ET should spontaneously occurs even at 0K
because the phonon bath can absorb efficiently the reaction energy. ET is then very fast.

When covalent interactions are present $ \Gamma_{xx}(0)\neq 0$,  the standard Marcus theory may drastically fail.
Then, either ET still occur but with an energy barrier which may be much smaller than those given by eq.\ref{enb} and may even vanish or a covalent bond forms instead of ET. 
The reason of this failure is that the standard Marcus theory only takes into account an overlap term which does not depend on the distance Donor Acceptor.
Actually it does because it is also obviously coupled to the phonon field. This covalent term may be often negligible so the standard Marcus theory is then successful. 
However, the covalent has no reason to be always negligible. We shall claim in a forthcoming paper that the existence of covalent term could be the key for a correct understanding of the so 
much puzzling enzymatic functions in biochemistry which involve soft molecules and electronic orbitals sometime quite extended.

Considering the limit  $\Gamma_{zz}(0) << \Gamma_{xx}(0) $ and $\epsilon_z <\Gamma_{xx}(0) $ (then $\Gamma_{xx}(0)>> |\Gamma_{xz}(0)|$), there are two energy minima on the sphere (see fig.\ref{figsur2} left) corresponding both to covalent bonds
where $\theta=0$mod $\pi$  and $I=I_c\approx \frac{\epsilon_z}{2\Gamma_{xx}(0)}$ with $|I_c|<1/2$ and  two maxima  on  $\theta=\pi/2$ mod $\pi$.
Actually, only the lowest minimum is physically acceptable for the covalent bond. In physical situations, the overlap term $\Lambda(\mathbf{R})$ is expected to exponentially vanish at large distance Donor-Acceptor, but due to the linear approximation on $\Lambda(\mathbf{R})$ in (\ref{hamspin}), it does not and change sign. Thus potential $H_{eff}(I,\theta)$ is well-described only on the part of the sphere where $\Lambda(\mathbf{R})$ keeps the same sign as $\Lambda(\mathbf{0})$ and only the trajectories restricted to this region
are physically acceptable.  The same problem appears in the well-known SSH model \citet{footnote3}.
The poles $I=\pm 1/2$ are unstable because they belong to large amplitude time periodic orbits with frequency in the range of phonon frequencies. These trajectories are dissipative and converge toward  the minimum  energy solution which is the covalent bond. This is the situation of free radicals which spontaneously bind without activation energy. 

The most interesting situation is obtained in the intermediate case, when  both charge and covalent interactions are present.
Fig.\ref{figsur2} shows the ideal case obtained for well chosen parameters where $\epsilon_x=\epsilon_z=0$, $\Gamma_{xx}(0)=\Gamma_{zz}(0)$,
$\Gamma_{xz}(0)=0$. Then $H_{eff}(I,\theta)$ is minimum along two degenerate paths $\theta=0$ or $\pi/2$ which is quite similar to those
of a dimer model with Targeted Energy Transfer (TET)  \citet{AKMT01} but as noted above,  only one of these paths is physically relevant.
For model parameters near but not equal to their ideal values,  the energy profile between Donor and Acceptor is still rather flat 
with small energy barrier if any. Instead of pure degeneracy, there is a strong softening of the global dynamics
(involving both the electron and phonons motions). ET may then occur spontaneously and very fast at 0K 
providing the reaction energy be slightly positive without energy barrier.
If there is a small energy barrier, then ET would occur efficiently at a small temperature as soon the thermal energy is beyond this energy barrier.

We illustrate this dynamics for an example in this situation which could be understood independantly of the general theory.  Instead of  a direct numerical integration of eqs.\ref{neweq2}
for a model involving a phonon continuum,
it is  much easier to consider an equivalent model  where there is only a small number of phonons submitted to a standard damping and  a Langevin force with a white spectrum corresponding to a given bath at some temperature. We choose Hamiltonian
\begin{equation}
H= (\epsilon_z+ k_z u_z ) Z+ (\epsilon_x+ k_x u_x ) X +\frac{1}{2} p_z^2 + \frac{1}{2} \Omega_z^2 u_z^2 +\frac{1}{2} p_x^2 + \frac{1}{2} \Omega_x^2 u_x^2
\label{dimham}
\end{equation}
which involves two harmonic oscillators with unit mass and coordinates $u_z$ and $u_x$ with damping constants $\gamma_z$ and $\gamma_x$ respectively.
The first oscillator is only coupled to the charge term $Z=|\varphi_D|^2 -|\varphi_A|^2$  and the second one only to covalent term $X=\varphi_A\varphi_D^{\star}+\varphi_D\varphi_A^{\star}$.
Its dynamical equations are  
\begin{eqnarray}
 i \hbar \dot{\varphi}_D&=& (\epsilon_z +k_z u_z) \varphi_D +(\epsilon_x +k_x u_x) \varphi_A  \\
i \hbar \dot{\varphi}_A&=& - (\epsilon_z +k_z u_z) \varphi_A +(\epsilon_x +k_x u_x) \varphi_D  \\
 \ddot{u}_z &+& \gamma_z \dot{u}_z + \Omega_z^2 u_z + k_z Z=\eta_z(t) \\
 \ddot{u}_x &+&  \gamma_x \dot{u}_x + \Omega_x^2 u_x + k_x X=\eta_x(t)
 \label{coupeq}
 \end{eqnarray}
where  $\eta_z(t)$ (resp. $\eta_x(t)$ are random gaussian white noise at temperature $T$ which fulfills the Langevin condition
$<\eta_z(t+\tau)\eta_z(t)> = 2 k_B T \gamma_z \delta(\tau)$ , $<\eta_z(t+\tau)\eta_x(t)> =0$ and $<\eta_x(t+\tau)\eta_x(t)> = 2 k_B T \gamma_x \delta(\tau)$.
The harmonic oscillators variables may be eliminated in a similar way as in the general case (\ref{hamtot4}) which yields eqs.\ref{neweq2}  
with  memory kernels  with the form
\begin{eqnarray}
\Gamma_{zz}(t)& =& \frac{k_z^2}{\Omega_z^2} 
e^{-\frac{\gamma_z}{2}t} \left( \cos  \tilde{\Omega}_z t + \frac{\gamma_z}{2\tilde{\Omega}_z}  \sin  \tilde{\Omega}_z t \right)\\
 \Gamma_{xx}(t) &=& \frac{k_x^2}{\Omega_x^2} 
e^{-\frac{\gamma_x}{2}t} \left( \cos  \tilde{\Omega}_x t + \frac{\gamma_x}{2\tilde{\Omega}_x}  \sin  \tilde{\Omega}_x t \right)
\label{newparam}
\end{eqnarray}
Frequencies  $\tilde{\Omega}_z= \sqrt{\Omega_z^2 -\frac{\gamma_z^2}{4}}$ and  $\tilde{\Omega}_x= \sqrt{\Omega_x^2 -\frac{\gamma_x^2}{4}}$ may be real (underdamped case) or purely imaginary (overdamped case)
but in both case, the memory kernels remain real. There is no cross term  $\Gamma_{xz}(t) =0$ because phonons are either coupled to 
the charge term or to the covalent term but not to both. These functions at time $0$ determine the effective Hamiltonian (\ref{generham}) appearing in (\ref{neweq2}).
Forces $\zeta_z(t)$ and $\zeta_x(t)$ are random, gaussian and its correlations  obey eqs.\ref{Langevinrel}.

This modeling may be qualitatively correct only when the electronic frequencies are comparable with the phonon frequencies. The reason is that the memory functions of this model do not exibit a sharp cut-off at any $\omega_c$ and consequently electronic damping would persist in the BO regime while it should not. 

\begin{figure} 
    \centering
      \includegraphics[width= 
      0.8\textwidth]{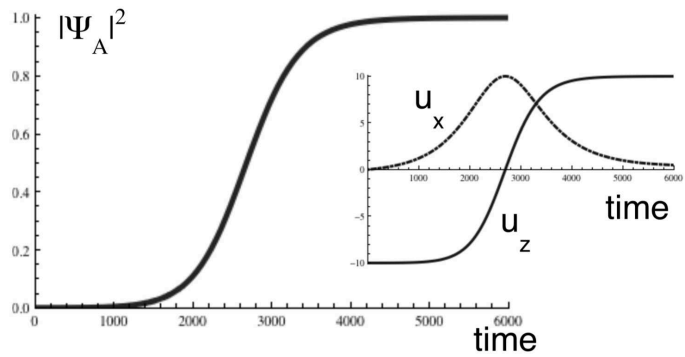}
    \caption{Electron Density on Acceptor versus time in the Dimer  example (\ref{coupeq}) (with $\hbar=1$) 
    at 0K. Insert shows the variation of  $u_z(t)$ ( representing the local reorganisation due to ET)
   while the covalent coordinate  $u_x(t)$ simultaneously varies only back and forth. Parameters are chosen to be in an almost isoenergetic situation $\epsilon_z=0.03,\epsilon_x=-0.001, k_z=k_x=0.1, \Omega_z=\Omega_x=0.1,\gamma_z=\gamma_x=0.2$.}
 \label{figdim}
\end{figure} 

We choose to illustrate our present study by an example at 0K but with ultrafast ET. 
We just integrate numerically  eqs.\ref{coupeq} at 0K (without random force) 
 which requires only few seconds of computer time using for example the programming language Mathematica.

Fig.\ref{figdim} shows an example of ultrafast ET versus time at 0K and the corresponding variations of the phonon variables.
 For this example, parameters were chosen in order to be in the barrierless situation (where $\Gamma_{zz}(0)=\Gamma_{xx}(0)$ and $\Gamma_{xz}(0)=0$)
and also to have  a small reaction energy  $\Delta G_0= 2\epsilon_z \approx 0.06 eV$ positive ( but not zero) (see fig.\ref{fig1}). 
Note that in that situation, the standard Marcus model would expect a relatively large energy barrier  $\Delta G^{\star} = \frac{\epsilon_z-\Gamma_{zz}(0))^2}{2 \Gamma_{zz}(0)}
\approx \frac{\Gamma_{zz}(0)}{2} \approx 0.5 eV$ which would prevent any efficient electron transfer up to room temperature.

To give some  idea of the time scale of ET predicted by our model  in physical situations,  reasonable  energy units should scale about the order of $1 eV=1.6 10^{-19} J$
while phonon energy quanta $\hbar \Omega_z$ and $\hbar \Omega_x$ should scale at most in the range of optical phonons energy $\sim 0.1 eV$.  Then, if we rescale the time unit in order that $\hbar=1$, the unit of time appearing fig.\ref{figdim} would range about $6.58 \times 10^{-16}$ second.Thus in this example the time for ET, would range physically about the order of
$1 ps$  which is ultrafast  (that is faster by many orders of magnitude than the characteristic time of  most electron transfer). Such ultrafast ET is possible even at 0K
but in rather optimized  regimes between underdamped and overdamped.

Actually, in our semiclassical theory, phonon damping do not matter for determining the energy profile and the energy barriers
but  it is essential for the dynamics of the electron transfer to be fast. 
Fastness at 0K requires of course a barrierless situation but also to be in a non adiabatic regime 
in order  the reaction energy may be quickly dissipated into the phonon bath. 
On contrary,  in adiabatic regimes even with no energy barrier, for example in the Marcus
inverted regime far from the inversion point, the electron dynamics is much faster than the phonon dynamics. 
There is no energy dissipation and ET cannot occur at 0K 
\footnote{unless by chemiluminescence that is very slowly by photon emission. Note that as well as the phonon bath, the photon bath is also  coupled to the ECR  and may also contribute to dissipate the reaction energy. But there are two  important differences. First the spectrum of the photon bath extends to infinity without any frequency cut-off (like $\omega_c$) and second the photon bath is generally only weakly coupled  to ECR in a regime where the above mean field (or semiclassical) approximation is not valid. Then, this coupling is usually treated as a quantum perturbation leading to a Fermi Golden rule for the transition probabilities.}.

A detailed study of ET as a function of temperature in our model  would be quite instructive for checking 
both Arrhenius and non Arrhenius behaviors in various regimes.  We have not done it. Actually,
this work would require much longer numerical calculations because of long statistics on the random forces. 
It is left for future work \footnote{Note that since eqs. \ref{coupeq} contain purely white random forces, it is possible to derive exact Fokker-Planck equations describing the time evolution of the probability density $P(\varphi_D, \varphi_A, u_z,\dot{u}_z, u_x,\dot{u}_x;t)$ in the phase space. However, 
it is not clear whether these equations would provide more efficient numerical methods (compared to a direct statistical study of the initial equations with random Langevin forces).}.

We expect an Arrhenius behavior when there is a large energy barrier or/and no phonon dissipation. Then, since ET at 0K is not possible, thermal fluctuations
are necessary for allowing ET. They generate random forces which acts adiabatically on the electronic state ( represented on the sphere $I,\theta$) and consequently its random diffusion
on this sphere.  Thus this electronic state may reach the vicinity of the transition state with activation energy  $\Delta G^{\star}$ (with the Arrhenius probability
proportional to $e^{-\frac{\Delta G^{\star}}{k_BT}}$). Then non-adiabatic effects can take place and produce ET (quantum tunneling) with some extra probability contributing to the prefactor $A(T)$.

If $\Delta G^{\star}$ becomes small or negligible,   the Arrhenius factor $e^{-\frac{\Delta G^{\star}}{k_BT}}$ becomes constant and unity in most range of temperature
so that  the reaction rate  cannot  obey anymore an Arrhenius law and is essentially described by its prefactor $A(T)$. 
As shown in our above example, when $\Delta G^{\star}=0$, ET occurs very fast at 0K  following a non random coherent trajectory. 
At non vanishing temperature, thermal fluctuations are expected to disturb the coherence of this ideal trajectory which should become stochastic.
One should expect a decrease of the rate of ET as  temperature grows. These intuitive predictions  could be numerically and quantitatively examined on our model (\ref{coupeq}).

In summary, we have  built a simple semiclassical theory of ECR using the complex electronic amplitudes as reaction coordinates instead of the nuclei coordinates.
This new formalism allows one to treat within the same model both charge and covalent interactions. In the limit where only charge interactions are present,  we recover the standard redox theory of ET (with extra refinements for describing the quantum tunneling without any empirical use of the Landau-Zeener effect). 
Our model may also be applied in situations where the Arrhenius law does not hold. With only the covalent interactions, we can model the covalent binding of free radicals (with no energy barriers). We also expect intermediate situations with finely tuned  charge and covalent interactions, with almost flat energy profile.
Those ECR are ultrafast elementary chemical reactions even at zero temperature.

I acknowledge George Kopidakis and Jos\'e Teixeira for valuable discussions and Laboratoire L\'eon Brillouin for its hospitality.

\end{document}